\documentclass[conference]{IEEEtran}
\IEEEoverridecommandlockouts
\usepackage{cite}
\usepackage{amsmath,amssymb,amsfonts}
\usepackage{algorithmic}
\usepackage{graphicx}
\usepackage{textcomp}
\usepackage{xcolor}

\usepackage{adjustbox}
\usepackage{tabularx}
\usepackage{listings}
\usepackage{hyperref}

\def\BibTeX{{\rm B\kern-.05em{\sc i\kern-.025em b}\kern-.08em
    T\kern-.1667em\lower.7ex\hbox{E}\kern-.125emX}}
\begin{document}

\title{Implementation of a 8-bit Wallace Tree Multiplier\\
}

\author{\IEEEauthorblockN{Ayan Biswas\textsuperscript{\textdagger}, Jimmy Jin\textsuperscript{\textdaggerdbl}}
\IEEEauthorblockA{\textit{Elmore Family School of Electrical and Computer Engineering, Purdue University} \\
West Lafayette, Indiana, USA \\
\textsuperscript{\textdagger} Doctor of Philosophy (PhD) Student, \textsuperscript{\textdaggerdbl} Master of Science (MS) Student \\
\{ayanb, jin357\}@purdue.edu}
}

\maketitle

\begin{abstract}
Wallace tree multipliers are a parallel digital multiplier architecture designed to minimize the worst-case time complexity of the circuit depth relative to the input size \cite{wallacetree}. In particular, it seeks to perform long multiplication in the binary sense, reducing as many partial products per stage as possible through full and half adders circuits, achieving \(\mathbf{O(log(n))}\) where \(\mathbf{n = }\textbf{ bit length of input}\).
This paper provides an overview of the design, progress and methodology in the final project of ECE 55900, consisting of the schematic and layout of a Wallace tree 8-bit input multiplier on the gpdk45 technology in Cadence Virtuoso, as well as any design attempts prior to the final product. This also includes our endeavors in designing the final MAC (Multiply Accumulate) unit with undefined targets, which we chose to implement as a 16 bit combinational multiply-add.
\end{abstract}

\section{Introduction}
\label{sec:intro}
\subsection{Long Multiplication}
 A key property of multiplication lies in its distributive property of multiplication over addition, namely, \(a \times (b + c) = a \times b + a \times c\). Long multiplication is a particular multiplication algorithm which exploits this, using the radix and the place value of the digit \cite{west2011}. This property can be applied irrespective of the radix of the system.
\subsection{Combinational Circuits}
A digital circuit takes in digital input signals - a mapping of a range of voltages \(f : [V_{SS}, V_{DD}] \rightarrow \{0, x, 1\}
\) \cite{kumar2016}. In our case, we use 1 V complementary metal oxide semiconductor (CMOS) technology. As such, with respect to ground, voltages sufficiently close to 1 V, or higher are represented digitally as logical 1, while voltages sufficiently close to 0 V, or lower are represented digitally as a logical 0. Anything in-between, we consider as x, or metastability. Combinational circuits are such digital circuits which take in only currently input values for the production of output values.
\subsection{Wallace Tree Multiplier}
A Wallace multiplier is a digital multiplier that takes in the inputs encoded in binary and produces another output encoded in binary. Our implementation is a combinational circuit which performs multiplication. Taking inspiration from long multiplication, a parallel multiplier uses the distributive property of multiplication over addition with the radix of 2. Long multiplication can be through of two components -- a part where all the intermediate products are determined, and another part where the final product is determined from the sum of all the intermediate products. 
\begin{figure}[h]
    \centering
    \includegraphics[width=0.5\linewidth]{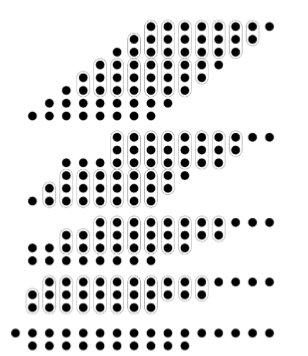}
    \caption{8-bit Tree Reduction \cite{wallaceimg}}
    \label{fig:treered}
\end{figure}
With Wallace multipliers, we consider the multiplier in whole as a 3 stage multiplier. First, intermediate products, or partial products, are produced through and-gates, such as \(a \land b = a \times b\) under \(GF(2)\). The final product is divided into two parts, the reduction part and the final addition part. An example of the reduction can be seen in Figure \ref{fig:treered}. The diagram is to be read from top to bottom, with each group showing the result after each reduction, starting from the output of the partial product generation. Note that groups of 3 are full adders, while groups of 2 are half adders, both producing a sum in the current column and a carry in the column to the left. Finally, once partial products are reduced to two rows of bits, we can use a simple digital adder to finish off the operation.

\subsection{Design Specifications}
For this design, we were tasked with designing the partial product generator, the full and half adder blocks, the adder arrays for each reduction pass, and the final adder (ripple-carry, carry select, etc) \cite{vijay2022}. 
\begin{itemize}
\item Input size 8
\item Input rise/fall time of 50 ps
\item Output load capacitance 2fF
\item DRC-clean and LVS-clean Layout
\item Reported worst case post-layout energy and propagation delay from extracted parasitics of layout
\item Identification of worst-case input(s) for delay with justification and corresponding metrics.
\end{itemize}

For evaluation, we must do the following.
\begin{itemize}
\item Pass functionality check
\item Worst-case propagation delay: \(\le\) 3ns
\item Worst-case input vector energy consumption: \(\le\) 1000fJ
\item Minimize area adhering to above performance constraints
\end{itemize}

\section{Architectural Exploration}
\label{sec:arch}
\begin{figure}[h]
    \centering
    \includegraphics[width=0.4\linewidth]{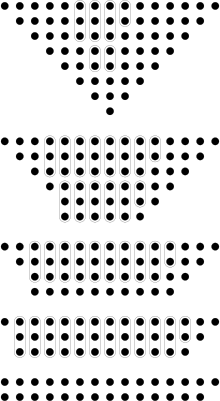}
    \caption{8-bit Dadda Tree Reduction \cite{daddaimg}}
    \label{fig:dadda_red}
\end{figure}

Prior to any circuit design based optimizations, such as choosing logic families, we first analyze the system architecturally for any architectural opportunities. The Wallace tree implementation allows for freedom of expression in methods of reduction, so long as the underlying methodology relies on the ability to reduce as much as possible with each row. In research, it is common to use an approach focusing on 4-2 compressions for lower accuracy 8-bit multipliers \cite{sureka2013}, as well as hybrid approaches such as combining Booth recoding with Wallace tree reductions \cite{rao2012}. The former was not chosen as a reduction of accuracy would fail functional tests in the constraints given, and the latter is not purely a Wallace tree implementation, and we do not do signed multiplication, rendering booth recoding useless. Other parallel reductions exists -- such as the Dadda multiplier, which aims to reduce the number of gates used while maintaining a similar reduction method as shown in Figure \ref{fig:dadda_red} \cite{dadda1965}.

Our initial approaches in architectural simplification aimed to reduce the area of our system. The intuition was that if we simplified the most commonly used primitive blocks -- full adders -- we would gain the most amount of increases when it comes to area. On initial evaluation, it was also obvious that the 3 ns threshold for latency would be easy to reach. Due to the composability of CMOS gates, an approximation can be seen from the total depth of logic gates. In the worst case, our critical path would be \(t_p = t_{p, and} + 4t_{p, FA} + 11t_{p,FA,RCA}\). Note that we chose to use gates due to the ease of layout compared to using full custom schematics for full adders and half adders.

\subsection{Alternate Compressors}
\begin{figure}[h]
    \centering
    \includegraphics[width=0.75\linewidth]{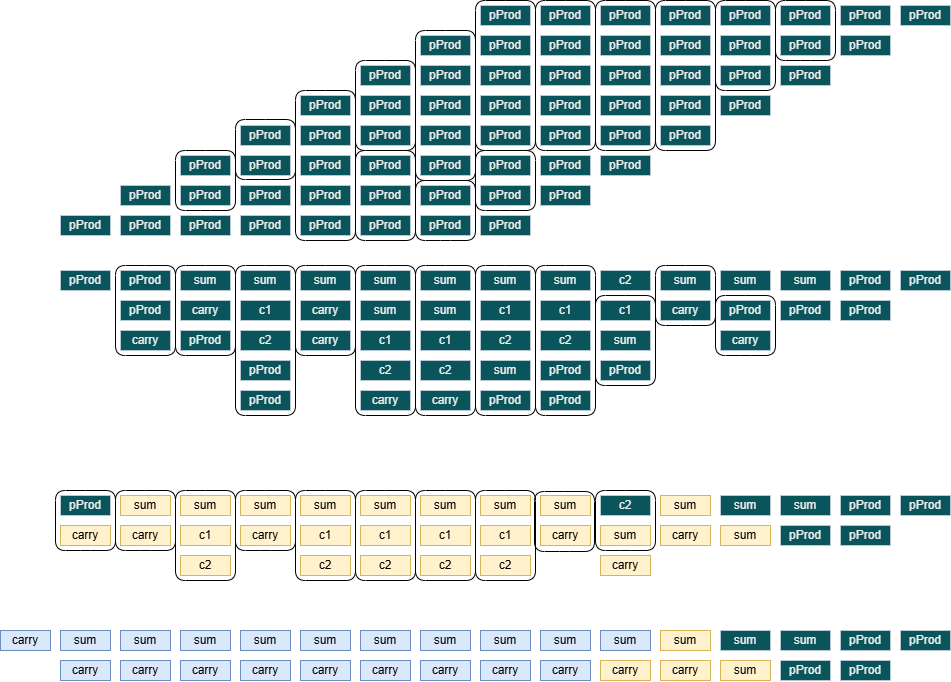}
    \caption{Sample Reduction using 5-3 Compressors}
    \label{fig:53sample}
\end{figure}

One key method which we could reduce area was to aggregate full-adders into modified primitive. 
\begin{figure}[h]
    \centering
    \includegraphics[width=\linewidth]{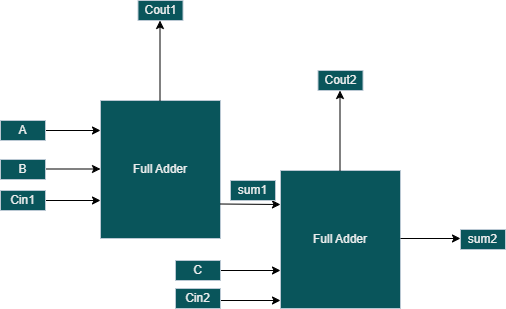}
    \caption{Complex Adder Composed of 2 Full Adders}
    \label{fig:CFA}
\end{figure}
A key advantage in this complex adder shown in Figure \ref{fig:CFA} lies in the reduction efficiency of such a unit. In essence, this is a non-lossy 4-2 compressor, or a 5-3 compressor. Each usage of such a unit reduces the number of signals at a ratio of \(\frac{5\text{ inputs}}{3\text{ outputs}}\), slightly better than a full adder at \(\frac{3\text{ inputs}}{2\text{ outputs}}\), and considerably better than a half adder at \(\frac{2\text{ inputs}}{2\text{ outputs}}\). This unit would be a fundamental trade-off between latency and area. By maximizing the usage of a such a unit, we can stay within bounds of the latency constraints and reduce both power and area by managing the sizes of these gates, as well as their fan-outs. An example of one such reduction is shown in Figure \ref{fig:53sample}.

Note that the reduction is slightly deceiving when it comes to latency. Although the new reduction is only 3 layers, it produces a slightly larger expected propagation delay for this stage due to complex adders taking roughly the delay of two full adders. Specifically, the new propagation delay is \(t_p = 2t_{p,CFA}+t_{p,FA} \approx 5t_{p,FA}\). Observe that the larger the proportion of the delay a part contributes to, the more effect does that have on the total speedup. Since the delay of the final adder, if implemented as a ripple carry adder is much larger, the increase in latency is fairly small. However, the final implementation does not use alternate compressors or an alternate reduction algorithm due to feedback from the Graduate Teaching Assistants.

\subsection{Negative Logic Units}
\begin{figure}[h]
    \centering
    \includegraphics[width=1\linewidth]{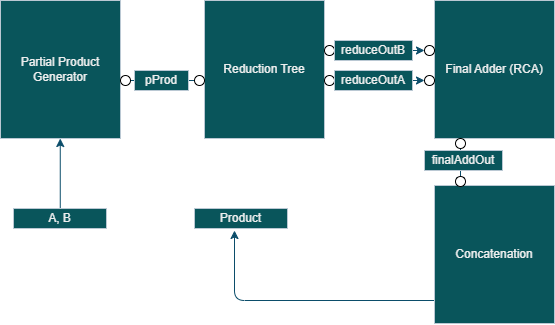}
    \caption{Negated Top Level RTL Diagram}
    \label{fig:negtoprtl}
\end{figure}

A key observation in the design is the pervasive usage of ANDs, ORs, and XORs, all of which are not natural to CMOS due to the inherent ability to create inverted logic blocks. For example, AND and OR are produced with a NAND or NOR in series with an inverter. XOR and XNOR are the exception to this as these gates are self-inverting, namely \(A \oplus B = \lnot A \oplus \lnot B\). Another key trait about static CMOS based design for XOR and XNOR lies in the usage of an extra inverter for the inputs, which could be shared with any other gates taht XOR or XNOR are used in. Hence, it makes sense to simplify full adders and half adders to use these gates in a negated fashion. The top level for such as design is shown in Figure \ref{fig:negtoprtl}.

Since adding a negation at the end of an AND is simply just a NAND, we reduce the system's inverter by 64, deferring the inversion until the final steps. Since each bit would need to be inverted at the end, this incurs an additional 16 inverters, providing a net benefit of 48 inverters. Taking the complement of the inputs and the complement of outputs for the boolean formulas for full adders and half adders lets us properly maintain the negation, as this allows the input to each stage to have an extra inverter, and the connections between each pair of adders to have 2 inverters logically, which by double negation results in the original input. Doing boolean algebra simplification yields the following for full adders and half adders.
\begin{align*}
    \text{sum}_{Complement~FA} &= \lnot(\lnot(A \oplus B) \oplus \text{cin})\\
    \text{carry}_{Complement~FA} &= \lnot\left(\lnot\left(\lnot(A \oplus B) \lor \text{cin}\right) \lor \lnot(A \lor B)\right)\\
    \text{sum}_{Complement~HA} &= \lnot \left( \lnot A \oplus \lnot B \right)\\
    \text{carry}_{Complement~HA} &= \lnot\left( \lnot A \land \lnot B \right)
\end{align*}

Note that the half adder has inverted inputs, but such a usage is fine due to the included inverters as part of the XOR. A schematic was made for this more optimized approach but is not included in the interest of space. The expected savings are evaluated based on summing up all the units and determining the difference in transistor counts for full adders, half adders, and the decreased usage of inverters. Notably, for a static cmos design using mux-based XORs and the traditional gate based implementation where the following are used for a guideline for 2-input gates:
\begin{align*}
    \text{sum}_{FA} &= \lnot(\lnot(A \oplus B) \oplus \text{cin})\\
    \text{carry}_{FA} &= ((A \oplus B) \land \text{cin}) \lor (A \land B)\\
    \text{sum}_{HA} &= A \oplus B\\
    \text{carry}_{HA} &= A \land B
\end{align*}
For the full adder, the original uses 2 12T XORs, 2 6T ANDs and a 6T OR, for a total of 42 transistors, while the converted uses 2 8T XORs, 3 4T NORs, and 4 inverters due to the lack of implicit inverters in XORS, for a total of 36 transistors. The half adder implementation uses 1 12T XOR and 1 6T AND for a total of 18 transistors, whereas the complement version uses 1 8T XNOR, 1 4T NAND, and 2 shared inverters for a total of 16 transistors. Since the reduction tree consists of 38 full adders and 16 half adders, and the final adder uses 11 full adders we save an estimated amount of \((38 + 11) \times 6 + 15 \times 2+48 = 372\) transistors with this approach. Further optimizations can be made by using a half adder rather than a full adder for the LSB of the final RCA. The final choice was to implement the non-negated version, but given more time, the negated version would be properly tested as well.

\subsection{Final Adder}
\begin{figure}[h]
    \centering
    \includegraphics[width=1\linewidth]{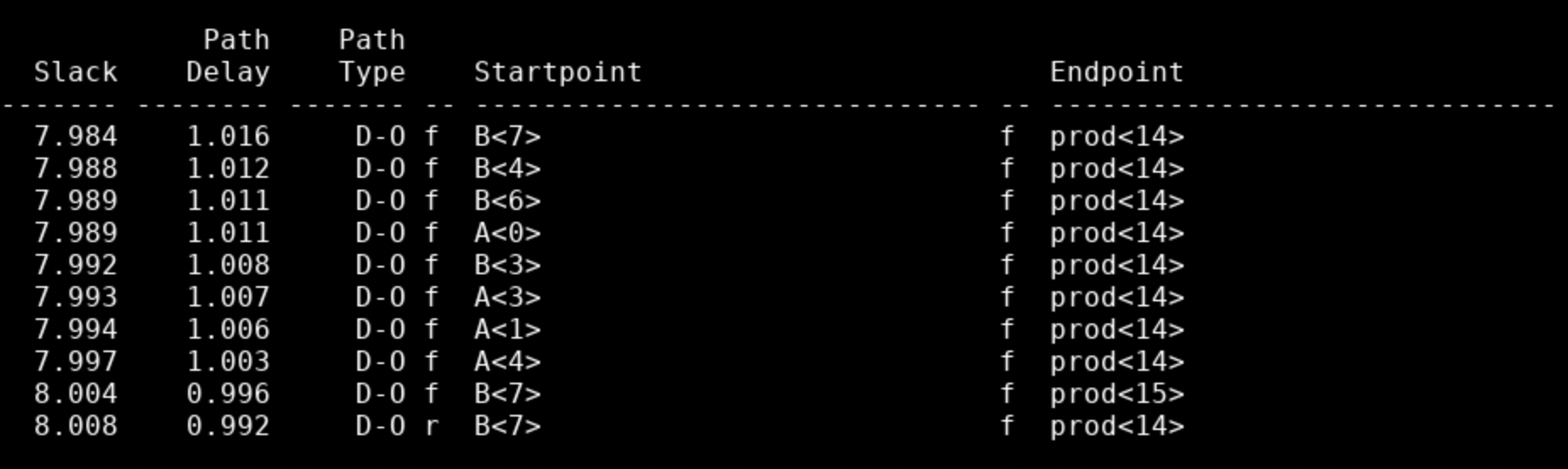}
    \caption{Nanotime for Path Delays of RCA Static CMOS Non-negated Logic}
    \label{fig:nanotimeworst}
\end{figure}

The choice of architecture for the final adder lies in the trade-offs of power, latency, and area. An RCA will have a critical path of 11 full adders, a carry select adder (CSA) will a critical path determined by the split -- assuming we use a sqrtCSA (square root carry select adder) with stages of 2-2-3-4, the critical path would be \(max(4t_{FA}, 3t_{FA} + t_{mux}, 2t_{FA} + 3t_{mux})\). Finally, we disregarded tree-based carry lookahead adders such as Kogge-Stone adders due to the wiring complexity and the fact that the \(O(log(n))\) complexity does not matter until sufficiently large bit lengths of adders \cite{chakali2013}. An adder of less than 16 bits does not tend to benefit a lot due to the need to go both up and down the tree and the difficulty in composing such an adder. Supposing the first term is maximal, a sqrtCSA reduces the expected critical path from \(t_{AND}+15t_{FA}\) to \(t_{AND}+9t_{FA}\), a respectable speedup, but suffers from greater wiring complexity, more than double the number of transistors for the final adder, the implementation of a multiplexer, only for an expected theoretical speedup of 50\%, not accounting for differences in extrinsic delays. We deemed such a speedup to be unnecessary due to our preliminary results on nanosim which yielded propagation delays of 1.1 ns or lower. An example for the RCA static CMOS non-negated logic implementation shows the following latencies in Figure \ref{fig:nanotimeworst}. The full results are stored in \$HOME/nanotime\_invert in our mg account (559mg14). Other simplifications can be done, such as observing that the MSB must not generate a carry, since \(8'b11111111 \times 8'b11111111 = 16'b1111111000000001\), meaning it could be a simple and between the MSB of the partial product and the overflow of a 10 bit RCA. Additionally, each start of a RCA inside a CSA could use an alternate half adder based on the input carry, as that value is predictable.

\section{Design of Multiplier}
\label{sec:design}
\subsection{RTL Diagram} 
\begin{figure}[h]
    \centering
    \includegraphics[width=1\linewidth]{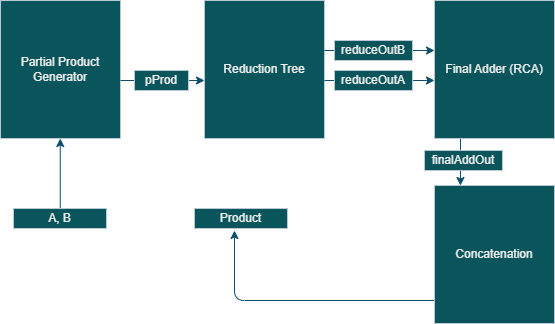}
    \caption{Top Level Multiplier RTL Diagram}
    \label{fig:toprtl}
\end{figure}

For flexibility in implementation, we created a diagram shown in Figure \ref{fig:toprtl}. The partial product generation is a simple array of and gates, 
\subsection{Schematic Design}

\subsubsection{Top Level Schematic}

\begin{figure}[h]
    \centering
    \includegraphics[width=\linewidth]{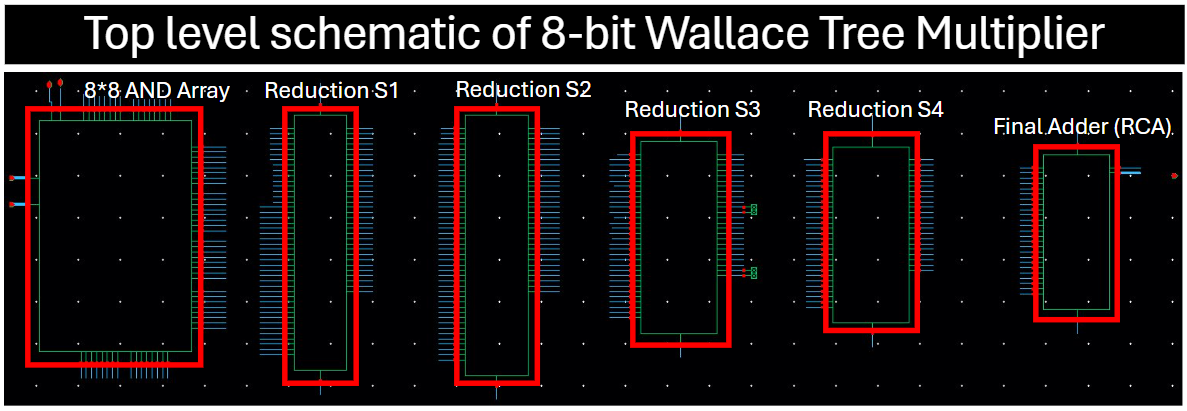}
    \caption{Top Level Schematic}
    \label{fig:top-schm}
\end{figure}

The schematic of an 8-bit Wallace Tree Multiplier, as shown in Fig. \ref{fig:top-schm}, demonstrates an efficient structure for multiplying two 8-bit numbers. The design begins with an 8×8 multiplier array, which generates partial products for all combinations of input bits. These partial products are then organized and reduced systematically using four stages of reduction (referred to as red stages) implemented with compressors or adders to minimize the number of partial product rows. Finally, the reduced partial products are summed up using a Ripple Carry Adder (RCA) to produce the final 16-bit product. This hierarchical approach reduces the critical path delay and improves the multiplier speed compared to conventional array multipliers.

\subsubsection{Partial Product Generator}

\begin{figure}[h]
    \centering
    \begin{adjustbox}{}
        \includegraphics[width=\linewidth]{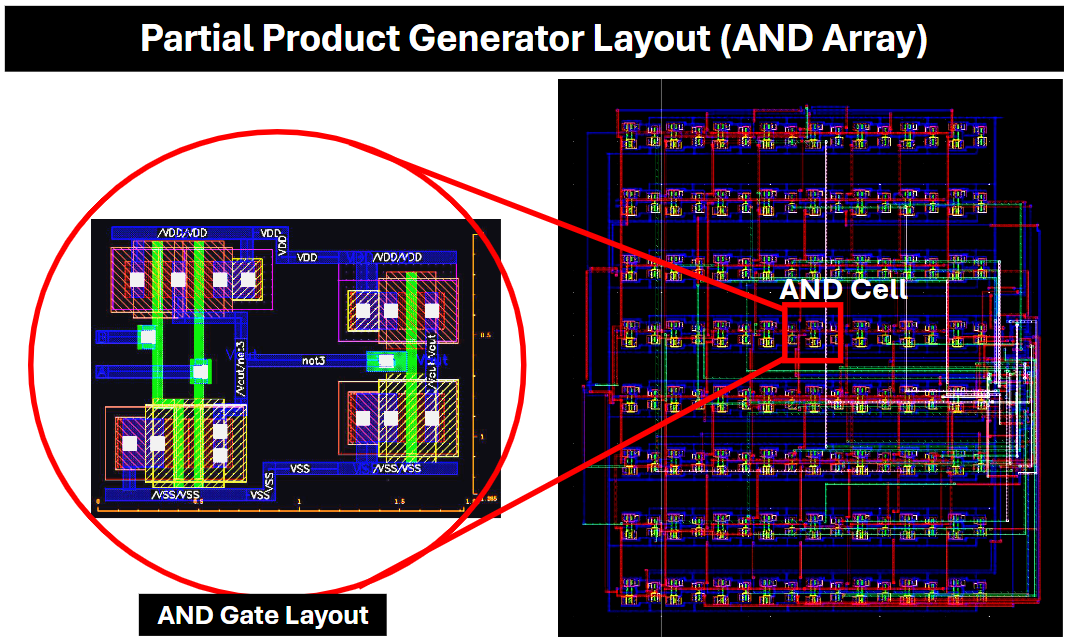}
    \end{adjustbox}
    \caption{Partial Product Generator}
    \label{fig:pp_gen}
\end{figure}

The partial product generator in the 8-bit Wallace Tree Multiplier consists of an array of 8×8 AND gates (as shown in Fig.~\ref{fig:pp_gen}), responsible for generating the initial partial products. Each AND gate computes the bitwise multiplication of one bit from the multiplicand and one bit from the multiplier, resulting in a total of 64 partial products. These partial products represent all possible combinations of input bits and form the basis for the subsequent reduction stages.

\subsubsection{Half Adder}

\begin{figure}[h]
    \centering
    \begin{adjustbox}{}
        \includegraphics[width=\linewidth]{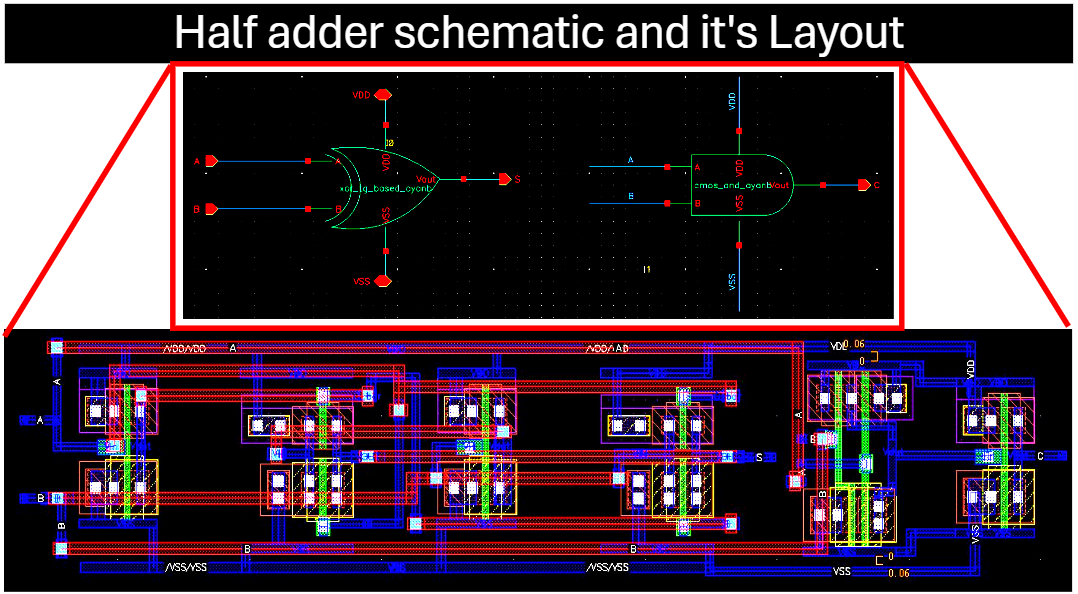}
    \end{adjustbox}
    \caption{Half Adder}
    \label{fig:ha}
\end{figure}

We have used a CMOS-based half adder (as shown in Fig.~\ref{fig:ha}) in the reduction stages of our Wallace Tree Multiplier to efficiently add multiple partial products. The half adder, implemented using CMOS technology, generates the sum and carry outputs for two single-bit inputs, contributing to the hierarchical reduction of partial product rows. Its low power consumption and high-speed operation make it well-suited for optimizing the performance of the multiplier.

\subsubsection{Full Adder}

\begin{figure}[h]
    \centering
    \begin{adjustbox}{}
        \includegraphics[width=\linewidth]{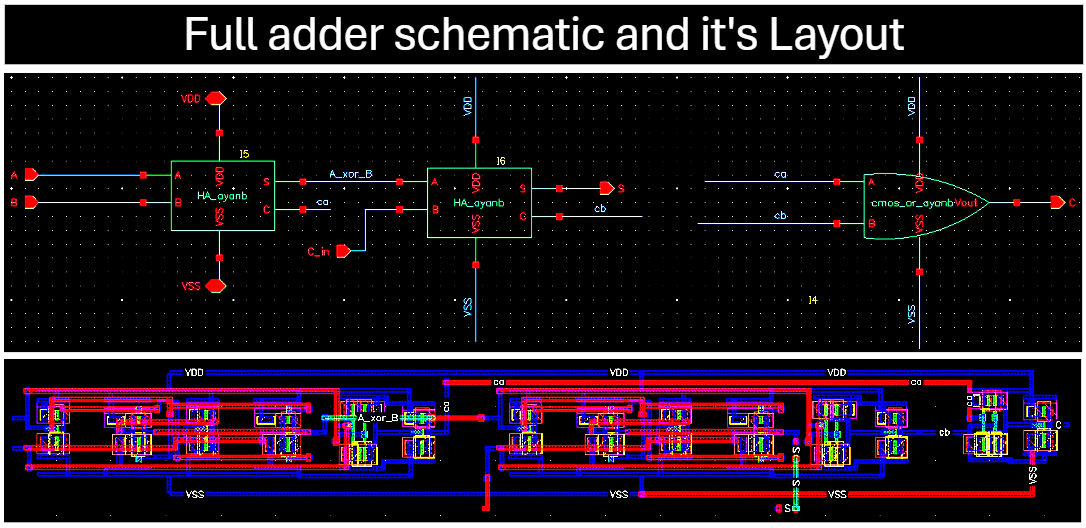}
    \end{adjustbox}
    \caption{Full Adder}
    \label{fig:fa}
\end{figure}

In the reduction stages of our Wallace Tree Multiplier, we have incorporated a CMOS-based full adder (as shown in Fig.~\ref{fig:fa}) to add multiple partial products. The full adder, built using CMOS technology, takes three inputs—two data bits and a carry input—and generates a sum and carry output. This functionality is essential for handling carry propagation and combining the partial products efficiently. By leveraging the low power consumption and fast operation of the CMOS full adder, we achieve improved performance in the multiplier's overall design.

\begin{figure}[h]
    \centering
    \begin{adjustbox}{}
        \includegraphics[width=\linewidth]{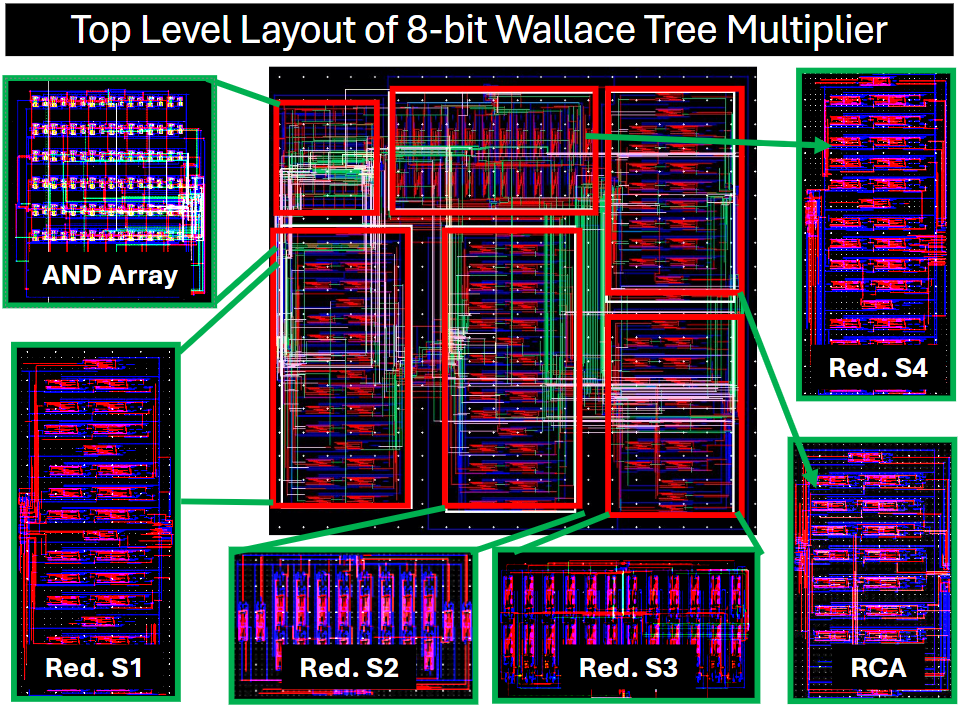}
    \end{adjustbox}
    \caption{top Level Layout}
    \label{fig:top_layout}
\end{figure}

\begin{figure}[h]
    \centering
    \begin{adjustbox}{}
        \includegraphics[width=\linewidth]{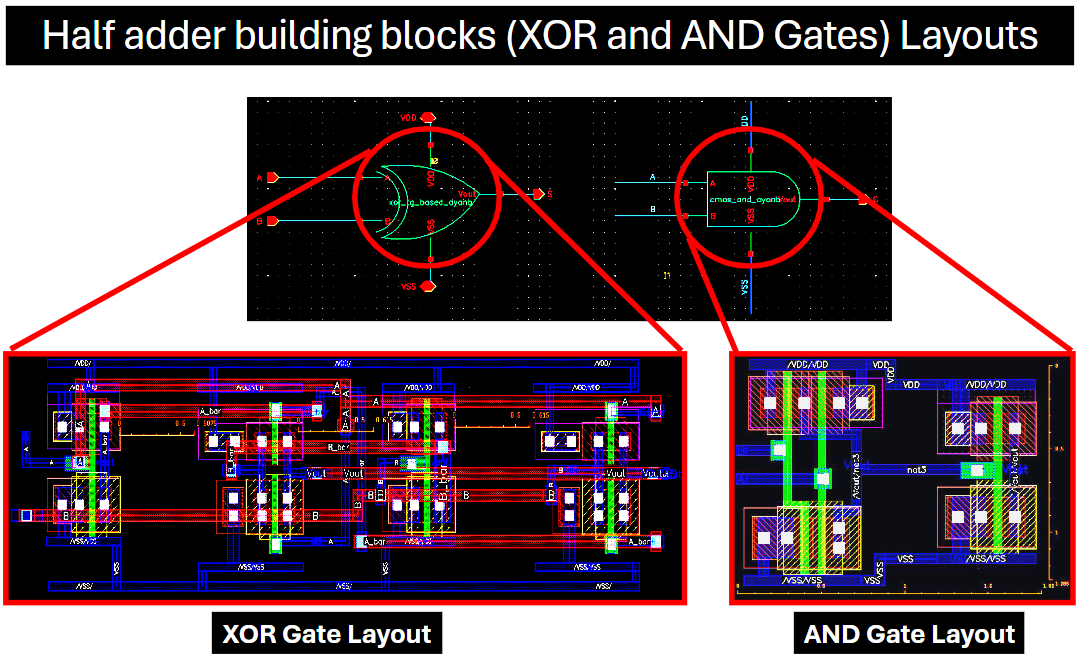}
    \end{adjustbox}
    \caption{XOR Gate and AND Gate Layouts}
    \label{fig:ha_build_layout}
\end{figure}

\begin{figure}[h]
    \centering
    \begin{adjustbox}{}
        \includegraphics[width=\linewidth]{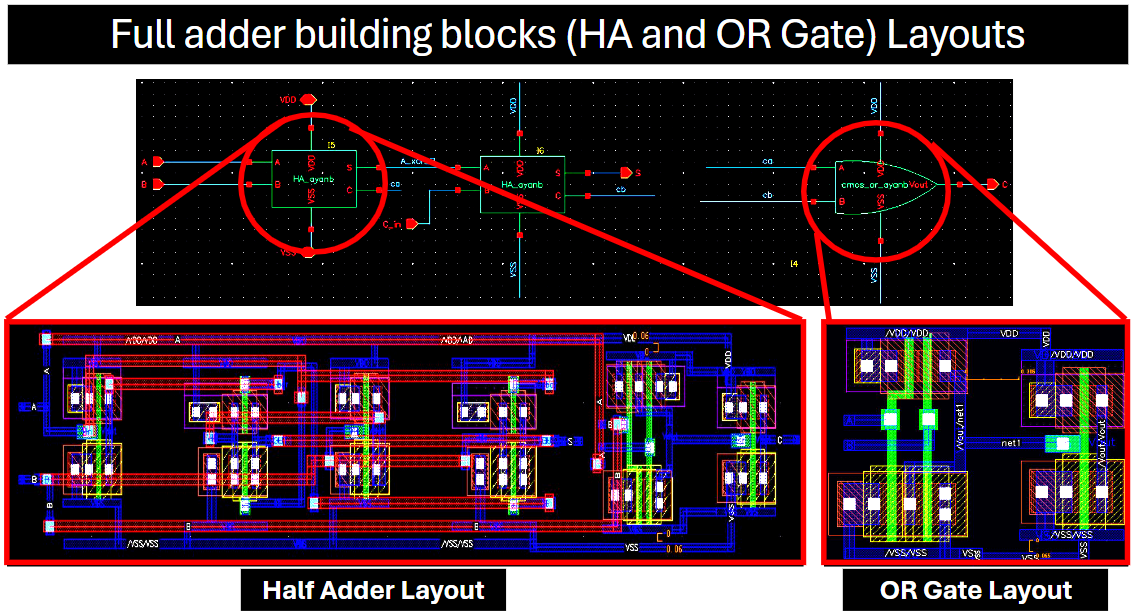}
    \end{adjustbox}
    \caption{Half Adder and OR Gate Layouts}
    \label{fig:fa_build_layout}
\end{figure}

\subsection{Topology Design for Circuit Primitives}

For our circuit primitives, we primarily utilized CMOS-based designs to ensure high noise immunity and low power consumption. However, for the XOR gate, we opted for a transmission gate (TG)-based implementation. While the transmission gate required a larger transistor width to offset delay, it significantly reduced the total transistor count compared to a traditional CMOS XOR gate. This reduction in the number of transistors proved advantageous in the top-level layout, conserving silicon area and simplifying the overall design without compromising functionality.

\subsection{Sizing Strategy}

In our sizing strategy, the transistors were dimensioned to equalize delays across the circuit by carefully adjusting their width (\(W\)) and length (\(L\)) relative to a reference inverter. The reference inverter serves as a baseline for delay, ensuring consistency in signal propagation. By scaling the transistor dimensions, we balanced the rise and fall times, minimizing any skew and optimizing the overall performance of the design. Larger \(W/L\) ratios were used for transistors driving higher capacitive loads to reduce delay, while smaller ratios were applied where minimal drive strength was sufficient. This approach ensured uniform delays and improved the timing performance of the circuit.

\subsection{DRC/LVS Reports}

\begin{figure}[ht]
    \centering
    \begin{adjustbox}{}
        \includegraphics[width=\linewidth]{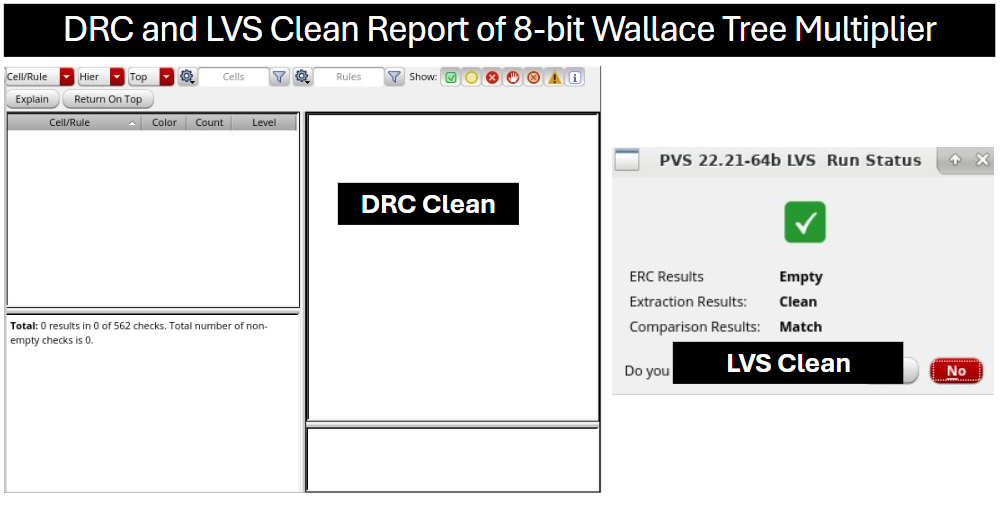}
    \end{adjustbox}
    \caption{DRC and LVS Cleaned layout of Multiplier}
    \label{fig:drc_lvs_mult}
\end{figure}

The entire system was designed in a modular manner, with each sub-block developed and verified independently to ensure design integrity. Before combining the sub-blocks into the top-level layout, all individual components were thoroughly checked and validated to be DRC (Design Rule Check) and LVS (Layout Versus Schematic) clean (as shown in Fig.\ref{fig:drc_lvs_mult}). This systematic approach ensured that the final top-level layout met all design rules and matched the intended schematic, minimizing errors and streamlining the verification process.

\subsection{Functionality}

\begin{figure}[h]
    \centering
    \begin{adjustbox}{}
        \includegraphics[width=0.95\linewidth]{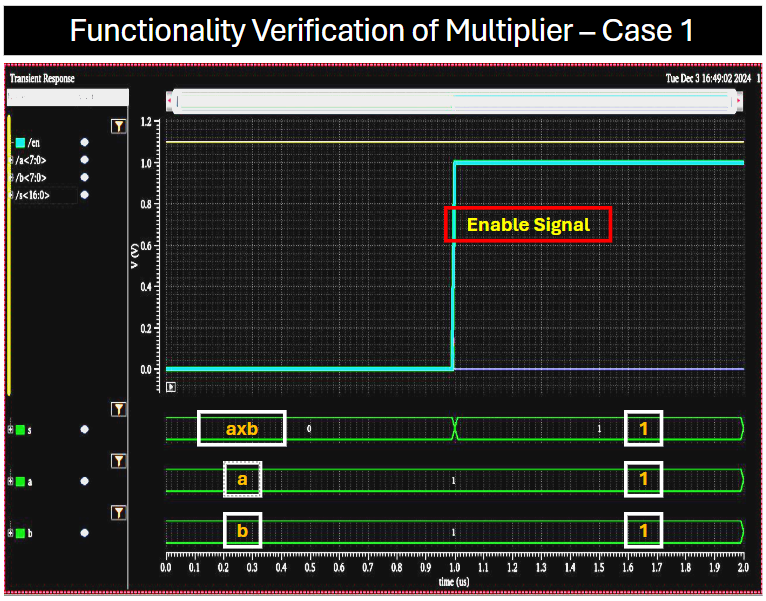}
    \end{adjustbox}
    \caption{Functionality Verification - Case 1}
    \label{fig:func1}
\end{figure}

\begin{figure}[h]
    \centering
    \begin{adjustbox}{}
        \includegraphics[width=0.95\linewidth]{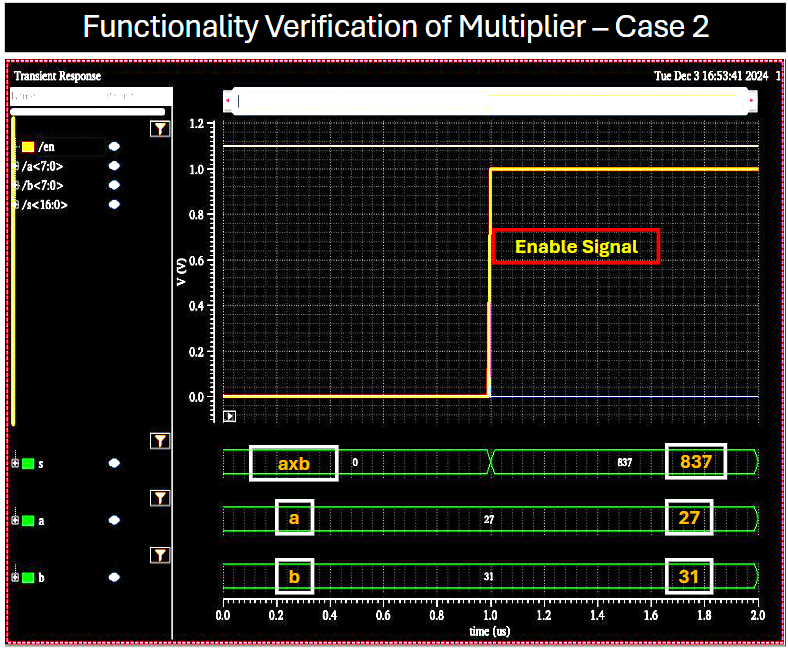}
    \end{adjustbox}
    \caption{Functionality Verification - Case 2}
    \label{fig:func2}
\end{figure}

\begin{figure}[h]
    \centering
    \begin{adjustbox}{}
        \includegraphics[width=0.95\linewidth]{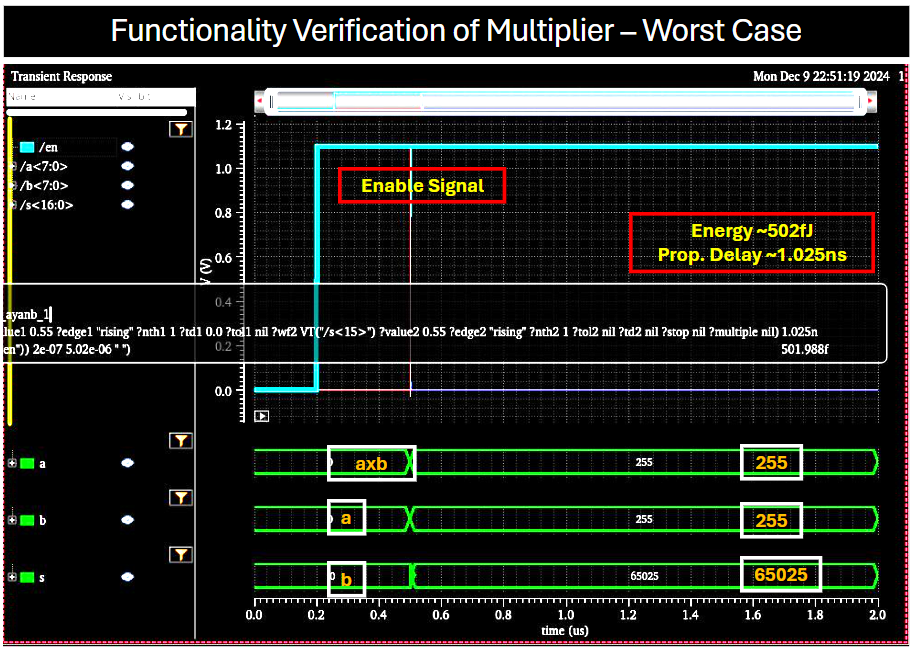}
    \end{adjustbox}
    \caption{Functionality Verification - Worst Case input pair}
    \label{fig:func4}
\end{figure}

The functionality of the Wallace Tree Multiplier was verified through simulation with various test cases, as shown in the following figures. In Figure \ref{fig:func3}, the result of multiplying 0 by 255 is shown, which correctly gives 0. Similarly, Figure \ref{fig:func1} demonstrates the multiplication of 1 with 1, yielding the expected result of 1. The case of multiplying 27 by 31 is shown in Figure \ref{fig:func2}, where the result is 837, as expected. Finally, Figure \ref{fig:func4} illustrates the worst-case input pair, obtained by looking for the longest carry chain in our multiplier, verifying the multiplier’s ability to handle extreme values. These results confirm the functional correctness of the multiplier design.

\begin{figure}[h]
    \centering
    \begin{adjustbox}{}
        \includegraphics[width=0.95\linewidth]{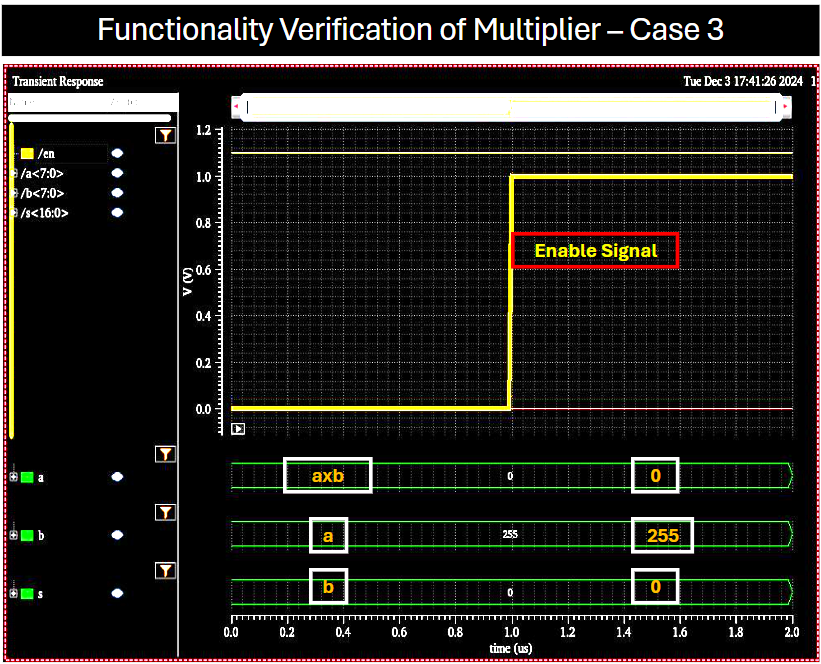}
    \end{adjustbox}
    \caption{Functionality Verification - Case 3}
    \label{fig:func3}
\end{figure}

\section{Energy-Latency-Area Analysis}
\label{sec:ener_lat}

\subsubsection{Gate Count for the Multiplier Design}
In Table \ref{tab:component_count}, we have presented the detailed gate count analysis for the Wallace Tree Multiplier. The table provides a breakdown of the number of Half Adders (HA), Full Adders (FA), AND gates, OR gates, and XOR gates used at each stage, including the 8×8 multiplier, reduction stages, and the final Ripple Carry Adder (RCA). In addition, it includes a summary of the total number of PMOS and NMOS transistors required for the design, highlighting the efficient use of resources. This analysis offers insight into the overall complexity and implementation at the transistor level of the multiplier, demonstrating its modular and optimized design.

\clearpage
\begin{table}[h]
\begin{center}
\caption{Component Count and MOS Cell Summary for Wallace Tree Multiplier}
\renewcommand{\arraystretch}{1.5} 
\setlength{\tabcolsep}{4pt} 
\begin{tabular}{|c|c|c||c|c|c||c|c|}
\hline
\textbf{Top Level Block} & \textbf{HA} & \textbf{FA} & \textbf{AND} & \textbf{OR} & \textbf{XOR} & \textbf{PMOS} & \textbf{NMOS} \\ \hline
8x8 AND Array       & -           & -           & 64           & -           & -           & 192                & 192                \\ \hline
Reduction S1             & 4           & 12          & 28           & 12          & 28          & 232                & 232                \\ \hline
Reduction S2             & 3           & 13          & 29           & 13          & 29          & 242                & 242                \\ \hline
Reduction S3             & 4           & 8           & 20           & 8           & 20          & 164                & 164                \\ \hline
Reduction S4             & 4           & 7           & 18           & 7           & 18          & 147                & 147                \\ \hline
Final RCA        & 0           & 12          & 24           & 12          & 24          & 204                & 204                \\ \hline
\textbf{Total Gates} & -           & -           & 183          & 52          & 119         & \textbf{1181}      & \textbf{1181}      \\ \hline
\textbf{Total MOS} & -           & -           & -            & -           & -           & \multicolumn{2}{c|}{\textbf{2362}} \\ \hline
\end{tabular}
\label{tab:component_count}
\end{center}
\end{table}

\subsubsection{Energy-Latency-Area Measurements}

The energy, latency, and area analysis for the 8-bit Wallace Tree multiplier is presented in Table \ref{tab:energy_latency_area}.

\begin{table}[h]
\begin{center}
\caption{Energy, Latency, and Area Analysis for 8-bit Wallace Tree Multiplier}
\renewcommand{\arraystretch}{1.5} 
\setlength{\tabcolsep}{4pt} 
\begin{tabular}{|c|c|c|}
\hline
\textbf{Parameter} & \textbf{Schematic} & \textbf{Post-Layout} \\ \hline
\textbf{Energy (fJ)}         & 502   & 595   \\ \hline
\textbf{Propagation Delay (ns)} & 1.025 & 1.325 \\ \hline
\textbf{Multiplier Area}     & \multicolumn{2}{c|}{
    \begin{tabular}[c]{@{}c@{}}
    $97.53 \, \mu\text{m} \times 95.405 \, \mu\text{m}$ \\
    $= 9304.84 \, \mu\text{m}^2$
    \end{tabular}} \\ \hline
\end{tabular}
\label{tab:energy_latency_area}
\end{center}
\end{table}

\section{Bonus MAC unit}
\label{sec:mac}

\subsection{Design of the MAC Unit}
\begin{figure}[h]
    \centering
    \includegraphics[width=1\linewidth]{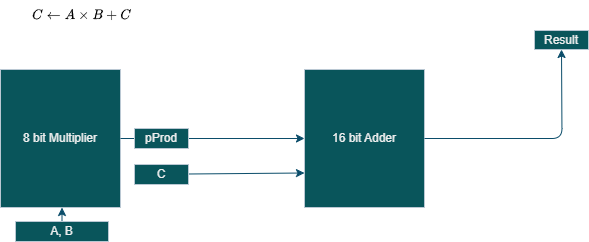}
    \caption{MAC Unit Top Level Diagram}
    \label{fig:mac_top_rtl}
\end{figure}

Figure \ref{fig:mac_top_rtl} shows our choice of design for our MAC unit. We do not choose to add a latch to feed in the 16 bit adder output back into C due to the lack of constraints, and thus made the assumption that we are building a simple fused multiply add. However, the addition of such elements would be trivial, including 16 flip flops and rewiring the output to C.

\begin{figure}[h]
    \centering
    \begin{adjustbox}{}
        \includegraphics[width=\linewidth]{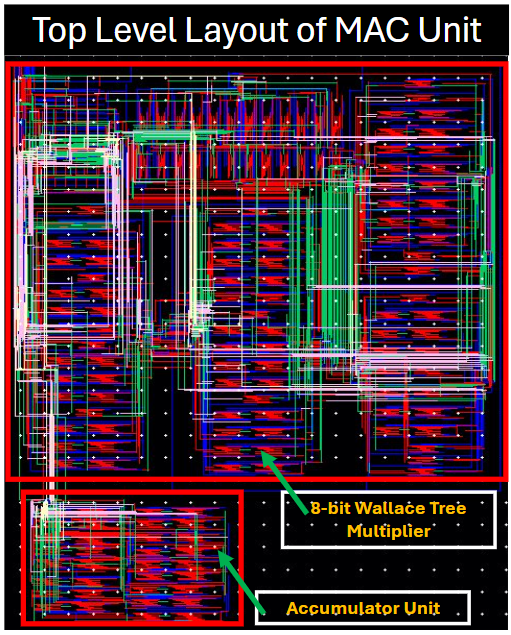}
    \end{adjustbox}
    \caption{Mac Unit Layout}
    \label{fig:mac_layout}
\end{figure}

\subsection{Functionality Verification}

\begin{figure}[h]
    \centering
    \begin{adjustbox}{}
        \includegraphics[width=\linewidth]{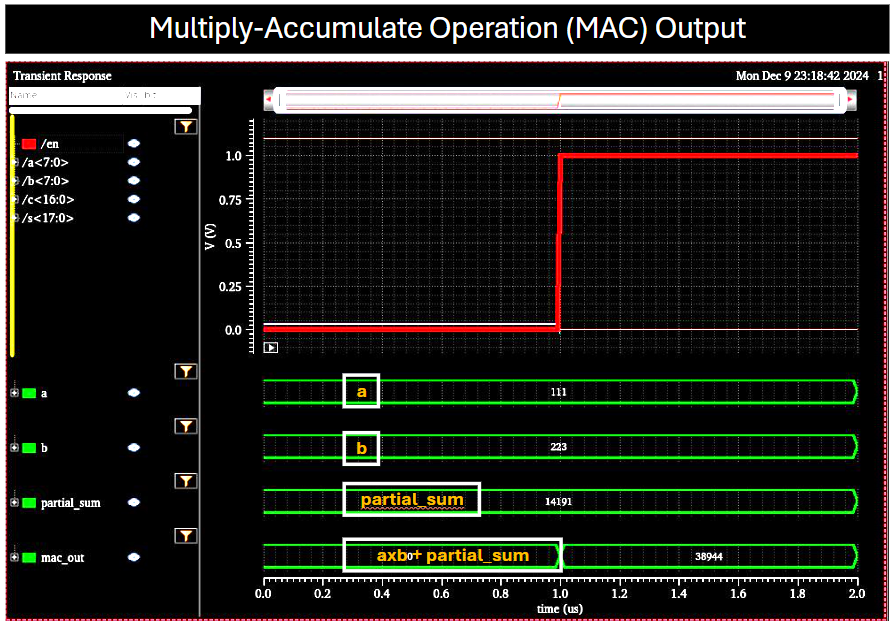}
    \end{adjustbox}
    \caption{Mac Unit Functionality Verification}
    \label{fig:mac_func}
\end{figure}

The functionality of the Multiply-Accumulate (MAC) unit was verified through simulation by testing specific input cases. As shown in Fig.\ref{fig:mac_func}, the MAC unit correctly computes the operation \(A \times B + \text{Partial Sum}\), where \(A = 11\), \(B = 223\), and the partial sum is 14191. The unit successfully calculates \(111 \times 223 + 14191 = 38944\), confirming its ability to perform accurate multiply-accumulate operations. This verification demonstrates the functional correctness of the MAC unit design.

\subsection{Energy-Latency-Area}

\subsubsection{Gate Count for the MAC Unit Design}
The total gate count and analysis of the number of PMOS and NMOS transistors for the MAC unit are presented in Table \ref{tab:mac_component_count}.

\begin{table}[h]
\begin{center}
\caption{Component Count and MOS Cell Summary for MAC Unit}
\renewcommand{\arraystretch}{1.5} 
\setlength{\tabcolsep}{4pt} 
\begin{tabular}{|c|c|c||c|c|c||c|c|}
\hline
\textbf{Top Level Block} & \textbf{HA} & \textbf{FA} & \textbf{AND} & \textbf{OR} & \textbf{XOR} & \textbf{PMOS} & \textbf{NMOS} \\ \hline
Accumulator Unit & 0           & 17          & 34           & 17          & 34          & 289                & 289                \\ \hline
8-bit Multiplier & -           & -           & 183          & 52          & 119         & 1181                & 1181                \\ \hline
\textbf{Total Gates}     & -           & -           & 217          & 69          & 153         & \textbf{1470}      & \textbf{1470}      \\ \hline
\textbf{Total MOS}       & -           & -           & -            & -           & -           & \multicolumn{2}{c|}{\textbf{2940}} \\ \hline
\end{tabular}
\label{tab:mac_component_count}
\end{center}
\end{table}

\subsubsection{Energy-Latency-Area Measurements}
The energy, latency, and area analysis for the MAC unit are presented in Table \ref{tab:energy_latency_area_mac}. Note that we used similar worst-case inputs of \(255 + 255 + 65535\) to maximize carry propagation stages.

\begin{table}[h]
\begin{center}
\caption{Energy, Latency, and Area Analysis for MAC Unit}
\renewcommand{\arraystretch}{1.5} 
\setlength{\tabcolsep}{8pt} 
\begin{tabular}{|c|c|c|}
\hline
\textbf{Parameter} & \textbf{Schematic} & \textbf{Post-Layout} \\ \hline
\textbf{Energy (fJ)}         & 582   & 620   \\ \hline
\textbf{Propagation Delay (ns)} & 1.152 & 1.412 \\ \hline
\textbf{Area}     & \multicolumn{2}{c|}{
    \begin{tabular}[c]{@{}c@{}}
    Multiplier: $9304.84 \, \mu\text{m}^2$ \\
    Accumulator: \\$38.815 \, \mu\text{m} \times 25.165 \, \mu\text{m} = 976.77 \, \mu\text{m}^2$ \\
    Total (MAC): $10281.54 \, \mu\text{m}^2$
    \end{tabular}} \\ \hline
\end{tabular}
\label{tab:energy_latency_area_mac}
\end{center}
\end{table}

\section{Conclusions}
\label{sec:conc}

In conclusion, this design project successfully involved the creation, layout, and verification of an 8-bit Wallace Tree multiplier along with a MAC unit. The multiplier was designed with a propagation delay of 1.325 ns, while the inclusion of the MAC unit resulted in a slightly higher delay of 1.412 ns. The total energy consumption for the multiplier was 595 fJ, whereas the MAC unit consumed 620 fJ, reflecting the added complexity. The total chip area for the multiplier was 9304.84 $\mu$m\(^2\), and for the complete MAC unit, it was 10,281.54 $\mu$m\(^2\). 

Through this project, we gained valuable experience in layout design and schematic development using Cadence Virtuoso, enhancing our understanding of both theoretical and practical aspects of circuit design. The project also provided insights into optimizing power, delay, and area, which are essential for efficient hardware design. Additionally, we learned the importance of modularity and the critical role of design verification in ensuring functional correctness. This project has significantly contributed to our knowledge of digital system design and its real-world applications.

\section{Contributions}
\label{sec:contri}
The layout and schematic for the final version multiplier and MAC unit, as well as generated images for Cadence Virtuoso were by Ayan Biswas. The analysis of various architectures, extra schematics for negated logic variants, Nanosim results, and verilog based testing were by Jimmy Jin. For the final paper submission, the abstract, the appendix, sections \ref{sec:intro}, \ref{sec:arch}, and \ref{sec:contri} were by Jimmy Jin; sections \ref{sec:ener_lat}, \ref{sec:mac}, and \ref{sec:conc} were by Ayan Biswas, and sections \ref{sec:design} and references were by both members of the team. Contributions are labeled according to the primary contributor.

\vspace{12pt}
\section*{Appendix}
\begin{itemize}
    \item \href{https://gist.github.com/ttlapik/c02b049754ae75a14a3ec740fbad9357}{\underline{559 Guideline on GitHub}}
    \item \href{https://drive.google.com/file/d/1BAAos22_KDvBHlnA5kcZ2Oluqfg8G3_H/view?usp=sharing}{\underline{Wallace Tree RTL Diagrams on draw.io}}
\end{itemize}
\end{document}